\documentclass[iop]{emulateapj-rtx4}
\usepackage{amsmath}
\usepackage{bm}
\usepackage{threeparttable}
\usepackage{CJK}
\def\bfnabla{{\mbox{\boldmath $\nabla$}}}

\renewcommand\bv{{\mbox{\boldmath $v$}}}
\newcommand\bb{{\mbox{\boldmath $B$}}}
\newcommand\bP{{\mbox{\boldmath $P$}}}
\newcommand\bn{{\mbox{\boldmath $n$}}}
\newcommand\bH{{\mbox{\boldmath $H$}}}

\newcommand\bF{{\mbox{\boldmath $F$}}}

\newcommand\Crat{{\mathbb{C}}}
\newcommand\Prat{{\mathbb{P}}}

\def\<{\,\langle\langle}
\def\>{\,\rangle\rangle}

\begin{document}
\begin{CJK*}{UTF8}{gbsn}

\shortauthors{Y.-F. Jiang et al.}

\author{Yan-Fei Jiang (姜燕飞)\altaffilmark{1\footnote{Einstein Fellow}}, James M. Stone\altaffilmark{2} \& Shane W. Davis\altaffilmark{3}}


\affil{$^1$Harvard-Smithsonian Center for Astrophysics, 60 Garden Street, Cambridge, MA 02138, USA} 
\affil{$^2$Department of Astrophysical Sciences, Princeton
University, Princeton, NJ 08544, USA} 
\affil{$^3$Canadian Institute for Theoretical Astrophysics. Toronto, ON M5S3H4, Canada}

\title{An Algorithm for Radiation Magnetohydrodynamics Based on
Solving the Time-dependent Transfer Equation}

\begin{abstract}
In order to incorporate the effects of radiation transfer in highly dynamical flows
where the light-crossing time is only marginally shorter than dynamical timescales, it
is necessary to solve the time-dependent transfer equation.  We describe 
a new algorithm for solving the coupled frequency-integrated transfer equation and
the equations of magnetohydrodynamics in this regime.
The transfer equation is solved in the mixed frame, including velocity dependent 
source terms accurate to $\mathcal{O}\left(v/c\right)$.  An operator split approach is
used to compute the specific intensity along discrete rays, with upwind monotonic
interpolation used along each 
ray to update the transport terms, and implicit methods used to compute the
scattering and absorption source terms.  Conservative differencing is used
for the transport terms, which ensures the 
specific intensity (as well as energy and momentum) are conserved along each ray to round-off error.
The use of implicit methods for the source terms
ensures the method is stable even if the source terms are very stiff.
To couple the solution of the transfer equation to the MHD algorithms
in the {\sc Athena} code, we perform direct
quadrature of the specific intensity over angles to compute the
energy and momentum source terms.
We present the results of a variety of tests of the method, such as calculating
the structure of a non-LTE atmosphere, an advective 
diffusion test, linear wave convergence tests, and the well-known shadow test.
We use new semi-analytic solutions for radiation modified shocks to 
demonstrate 
the ability of our algorithm to capture the effects of an anisotropic radiation field accurately.
Since the method uses explicit differencing of the spatial operators, it shows excellent
weak scaling on parallel computers.  The method is ideally suited for problems in which
characteristic velocities are non-relativistic, but still within a few percent or more
of the speed of light.  The method is an intermediate step towards algorithms for fully
relativistic flows.

\end{abstract}

\keywords{(magnetohydrodynamics:) MHD $-$ methods: numerical $-$  radiative transfer $-$ accretion disks}

\maketitle

\section{Introduction}
\label{sec:introduction}
The dynamical effects of radiation are important in a variety of astrophysical systems.
For example, in black hole accretion disks radiation
is not only an important cooling mechanism, but also is the dominant source of
pressure in the radiation-dominated 
regime.  The standard thin disk model \citep[][]{ShakuraSunyaev1973}, in which
heating and radiative cooling are balanced locally, 
is only applicable when the accretion rate is $\sim 0.01-0.3$ of the Eddington rate.   
At larger accretion rates, the disk is thought to become 
geometrically and optically thick \citep[][]{Abramowiczetal1988,AbramowiczFragile2011},
and radial
advection may also become an important cooling mechanism.
At both super-Eddington and strongly sub-Eddington accretion rates, 
strong outflows are likely to be produced 
\citep[][]{BlandfordBegelman1999,OhsugaMineshige2013}.
While much insight into the structure and
dynamics of accretion disks over this broad range
of parameters has been made using analytic steady-state models, more realistic 
calculations require numerical methods.

In addition, since the magneto-rotational instability (MRI)
\citep[][]{BalbusHawley1991,BalbusHawley1998} 
is generally believed to be the physical mechanism for angular momentum transport
in black hole accretion disks, numerical methods for magnetohydrodynamics (MHD) that
include the effect of radiation are crucial to enable study of the saturation
and nonlinear regime of the MRI in such disks.
Shearing box simulations of the MRI in the local patches of a
radiation pressure dominated accretion disk
have significantly improved our understanding of the dynamics in this regime
\citep[][]{Turneretal2003,Turner2004,Hiroseetal2006,Kroliketal2007,Blaesetal2011,Jiangetal2013b}. 
For example, in this regime disks are found to undego thermal runaway
in local shearing box simulations \citep[][]{ShakuraSunyaev1976,Jiangetal2013c}. 
However, understanding the saturation of thermal runaways and 
the implications for observations of radiation pressure dominated disks 
require global simulations.

In order to model the thermal properties of accretion disks over
a wide range of parameters accurately requires, 
algorithms for radiative transfer (RT) that satisfy several criteria.
First, the algorithm must be accurate in both optically thick (e.g. the mid-plane)
and optically thin (e.g. the photosphere) regions. 
Second, it must be accurate over a wide range of ratios of radiation to gas pressures,
from very small values (when the 
accretion rate is small) to very large values (when the accretion rate is high and the
disk is radiation pressure dominated).
Third, it must handle the non-local and anisotropic transport of photons accurately.
For example, the radiation field above the
photosphere will be produced by emission from many different (including distant)
parts of the disk.  This is likely critical to calculate the dynamics 
of outflows correctly. 
Finally, the algorithm must be efficient and highly parallelizable in order to enable
large-scale global simulations on modern computer systems.

Since in general the RT equation is a function of seven independent variables  
(time, spatial position, angles and frequency), it is usually thought that a formal
solution to compute the specific intensity is intractable, even in the case of
frequency independent opacities such that the grey (frequency independent) transport
equation can be used.
Thus, most algorithms for radiation MHD adopt a moment formalism, in which a
hierarchy of angle-integrated time-dependent moment equations are solved.  This
reduces the dimensionality of the problem to time and spatial coordinates, as are
used for the MHD equations.  However, the hierarchy of moment equations require
a closure relation.  A variety of different closures have been adopted, including the 
flux-limited diffusion (FLD) approximation 
\citep[][]{LevermorePomraning1981,TurnerStone2001,Krumholzetal2007,Zhangetal2011,Holstetal2011,Commerconetal2011} 
and more recently the M1 method
\citep[][]{Gonzalezetal2007,SkinnerOstriker2013,Sadowskietal2013,McKinneyetal2013}.
A variety of recent studies of the global dynamics of black hole 
accretion disks \citep[][]{Ohsugaetal2011,Sadowskietal2013,McKinneyetal2013} have
used such approximate closures.  
However, it is far from clear that such closures will be accurate everywhere in the flow.  The angular distribution of the radiation field can be arbitrarily complex, 
particularly in optically thin regions. But these schemes make the simple assumptions that the Eddington tensor can 
be uniquely determined by the local radiation energy density and flux. And, at present, there is no means of improving the approximation in these 
closures -- there is no resolution parameter that can be increased to assess convergence.
Thus, it is worthwhile to develop new algorithms
that are based on a direct solution of the RT equation.  A key message of our
work is that modern computer systems have advanced to the point that formal
solution of the six dimensional grey RT equation is now feasible at every time step
of a MHD simulation, and that approximate closures of unknown accuracy are no longer
necessary.

In this vein, we recently
have described an algorithm for radiation MHD that is based on the
method of short characteristics to solve the \emph{time independent} RT equation
in multidimensions
\citep[][]{Davisetal2012}, an approach that was also used in earlier work
\citep[][]{Stoneetal1992,HayesNorman2003,HubenyBurrows2007}.
At every time step, short characteristics are used to compute the specific
intensity, from which direct quadrature is used to compute
the components of the variable Eddington tensor (VET) which serves as the closure
relation in the radiation 
moment equations \citep[][thereafter JSD12]{Jiangetal2012}.  This method has been 
successfully used to study the dynamics of radiation-dominated accretion disks
using the local shearing box simulations \citep[][]{Jiangetal2013b,Jiangetal2013c}. 
The primary assumption underlying this method is that the light-crossing time is much 
shorter than characteristic dynamical times in the flow, so that solutions of the
time-independent RT equation can be used to compute the VET.  This is an excellent
approximation in most flows.  However, in the inner regions of
black hole accretion disks, dynamical times become comparable to the light crossing
time, and solution of the time-dependent RT is required.
The goal of this paper is to describe 
such an algorithm based on ray tracing methods.
Although our algorithm is motivated by the study of 
black hole accretion disks, it is also applicable to any 
system that has characteristic flow velocity which is not much smaller than
the speed of light. 

Of course, within a few gravitational radii of the black hole, the flow will be
fully relativistic, and both general relativistic MHD and RT are required to
describe the dynamics.  While GRMHD flows including radiation using FLD or the M1
closure have recently been reported \citep[][]{Fragileetal2012,Sadowskietal2013,McKinneyetal2013}, 
developing algorithms based on the formal
solution of the RT equation in GR is a formidable challenge.  Instead, in this
paper we describe an intermediate step.  That is, we develop ray tracing methods
for the formal solution of the time-dependent RT equation, but restrict ourselves
to non-relativistic flows.  Thus, there is a relatively narrow window of velocities
where our method is
applicable; since when $v \ll c$ the time step in our algorithm is too small for it
to be efficient, while for $v \approx c$ the flow is relativistic.
Nonetheless, there is an interesting range of radii in black hole accretion disks
where our method is applicable. We report results of simulations studying this region
elsewhere.  Moreover, we are now extending the method described in this paper
to full GR.

Another popular approach
to solve the time-dependent RT equation are
Monte Carlo methods \citep[][and references therein]{Whitney2011}. 
These methods are accurate, flexible and have almost perfect parallel scaling, and
therefore are a very attractive direction for the future. 
However, the intrinsic noise associated with Monte Carlo makes the method 
very expensive to model radiation pressure dominated flows accurately
\citep[][]{Davisetal2012}.
While it may be possible to reduce the noise and computational cost in Monte Carlo using
novel approaches \citep[][]{Yusef-Zadehetal1984,Densmoreetal2007,Steinackeretal2013},
in this work we instead adopt ray tracing methods.  Although ray tracing
methods are more complex to implement, they do not suffer from noise.

This paper is organized as follows. The equations we solve are given in Section \ref{sec:equation}. 
The basic angular discretization scheme is described in Section \ref{sec:angles}. 
A complete description of the algorithm are given in Section \ref{sec:algorithm} and we show 
test results in Section \ref{sec:test} to demonstrate the capability of the method.
In Section \ref{sec:performance}, 
we test the speed and parallel performance of the code. Finally, we summarize in Section \ref{sec:summary}.

\section{Equations}
 \label{sec:equation}
 
Following \cite{Lowrieetal1999} and JSD12, we solve the RT equation 
in the mixed frame: the specific intensity is measured in an Eulerian frame 
while the radiation-material interaction terms are computed in the co-moving frame
of the fluid.  This requires a Lorentz transformation of variables between frames.
For non-relativistic flows, keeping only terms to
$\mathcal{O}\left(v/c\right)$
\citep{MihalasKlein1982},
where $v$ is the flow velocity and $c$ is the speed of light, the RT equation is 
 \begin{eqnarray}
 \frac{\partial I}{\partial t}+c\bn\cdot\bfnabla I&=&c\sigma_a\left(\frac{a_rT^4}{4\pi}-I\right)+c\sigma_s(J-I) \nonumber \\
 &+&3\bn\cdot\bv\sigma_a\left(\frac{a_rT^4}{4\pi}-J\right)
 \nonumber \\
 &+&\bn\cdot\bv(\sigma_a+\sigma_s)\left(I+3J \right)
 -2\sigma_s\bv\cdot\bH \nonumber \\
 &-&(\sigma_a-\sigma_s)\frac{\bv\cdot\bv}{c}J- (\sigma_a-\sigma_s)\frac{\bv\cdot(\bv\cdot{\sf K})}{c}.\nonumber \\
 \label{RTequation0}
 \end{eqnarray}
Note that as discussed by \cite{Lowrieetal1999}, in order to keep the RT equation 
self-consistent in a moving medium, some $\mathcal{O}\left(v/c\right)^2$ 
terms need to be kept compared to 
equation (2.28) of 
\cite{MihalasKlein1982}.  These are the last two in equation (\ref{RTequation0}).
Here the frequency integrated specific intensity $I$ is a function of time, spatial positions and angles with unit vector $\bn$. 
Moments of the angular quadrature over all the solid angles $\Omega$
 are defined as
 \begin{eqnarray}
 J&\equiv& \int Id\Omega,\nonumber \\ 
 \bH&\equiv& \int \bn Id \Omega, \nonumber \\ 
 {\sf K}&\equiv& \int \bn\bn I d\Omega. 
 \label{integrateangle}
 \end{eqnarray}
 They are related to the commonly used radiation energy density $E_r$, radiation flux $\bF_r$ and radiation pressure ${\sf P_r}$  
via $E_r=4\pi J$, $\bF_r=4\pi c\bH$, and ${\sf P_r}=4\pi {\sf K}$.\footnote{The definitions here differ from equations (13) - (15) of 
\cite{Davisetal2012} by a factor 
of $c$.} Absorption and scattering opacities (attenuation coefficients) are $\sigma_a$ and $\sigma_s$ respectively, $a_r$ is the
 radiation constant and $T$ is the gas temperature.  
 The time and velocity dependent terms in equation (\ref{RTequation0}) change significantly the algorithm for solving
the RT equation compared with the short characteristic method described 
 in \cite{Davisetal2012}.
 
 In order to couple the radiation field to the gas, we need to take the angular moments 
 of equation (\ref{RTequation0}) to get the radiation energy ($S_r(E)$) and momentum ($\bm{ S_r}(\bP)$) source terms (these are
easily confirmed to be exactly the same as in \cite{Lowrieetal1999} and JSD12).
The zeroth and first moments of the right hand side of equation
(\ref{RTequation0}) multiplied by $4\pi/c$ are
\begin{eqnarray}
S_r(E)&=&\sigma_a\left(a_rT^4-E_r\right)\nonumber \\
&+&\left(\sigma_a-\sigma_s\right)\frac{\bv}{c^2}\cdot\left[
\bF_r-\left(\bv E_r+\bv\cdot{\sf P_r}\right)\right], \nonumber\\
\bm{ S_r}(\bP)&=&-\frac{\left(\sigma_s+\sigma_a\right)}{c}\left[
\bF_r-\left(\bv E_r+\bv\cdot{\sf P_r}\right)\right] \nonumber\\
&+&\frac{\bv}{c}\sigma_a\left(a_rT^4-E_r\right).
\end{eqnarray}

Then the ideal MHD equations with radiation energy and momentum source terms are \citep[][JSD12]{Stoneetal2008}
\begin{eqnarray}
\frac{\partial\rho}{\partial t}+\bfnabla\cdot(\rho \bv)&=&0, \nonumber \\
\frac{\partial( \rho\bv)}{\partial t}+\bfnabla\cdot({\rho \bv\bv-\bb\bb+{{\sf P}^{\ast}}}) &=&-\bm{ S_r}(\bP),\  \nonumber \\
\frac{\partial{E}}{\partial t}+\bfnabla\cdot\left[(E+P^{\ast})\bv-\bb(\bb\cdot\bv)\right]&=&-cS_r(E),  \nonumber \\
\frac{\partial\bb}{\partial t}-\bfnabla\times(\bv\times\bb)&=&0.
\label{MHDEquation0}
\end{eqnarray}
In the above equations, $\rho$ is density, ${\sf P}^{\ast}\equiv(P+B^2/2){\sf I}$ (with ${\sf I}$
the unit tensor), and the magnetic permeability $\mu=1$.  The total gas energy density is
\begin{eqnarray}
E=E_g+\frac{1}{2}\rho v^2+\frac{B^2}{2},
\end{eqnarray}
where $E_g$ is the internal gas energy density.   We adopt an equation of state
for an ideal gas with adiabatic index $\gamma$, thus
$E_g=P/(\gamma-1)$ for $\gamma\neq 1$ and $T=P/R_{\text{ideal}}\rho$, where
$R_{\text{ideal}}$ is the ideal gas constant.

In order to get a better insight on the relative importance of different terms in the radiation 
MHD equations, following \cite{Lowrieetal1999} and JSD12, we convert the above equations into dimensionless form by 
adopting two ratios $\Crat=c/a_0$ and $\Prat=a_rT_0^4/P_0$, where $a_0$, $T_0$ 
and $P_0$ are the characteristic values of velocity, temperature and gas pressure respectively. Therefore, 
$\Crat$ is the dimensionless speed of light while $\Prat$ is a measure of relative importance between radiation 
pressure and gas pressure when gas temperature is around $T_0$. Units for $E_r$, ${\sf P_r}$, $I$, $\bH$ and ${\sf K}$ 
are $a_rT_0^4$ while $\bF_r$ has units of $ca_rT_0^4$. Then the full dimensionless 
radiation MHD equations are

\begin{eqnarray}
 \frac{\partial I}{\partial t}+\Crat\bn\cdot\bfnabla I&=&\Crat\sigma_a\left(\frac{T^4}{4\pi}-I\right)+\Crat\sigma_s(J-I) \nonumber \\
 &+&3\bn\cdot\bv\sigma_a\left(\frac{T^4}{4\pi}-J\right)
 \nonumber \\
 &+&\bn\cdot\bv(\sigma_a+\sigma_s)\left(I+3J \right)
 -2\sigma_s\bv\cdot\bH \nonumber \\
 &-&(\sigma_a-\sigma_s)\frac{\bv\cdot\bv}{\Crat}J- (\sigma_a-\sigma_s)\frac{\bv\cdot(\bv\cdot{\sf K})}{\Crat}, \nonumber\\
 \label{RTequation}
\end{eqnarray}
\begin{eqnarray}
\frac{\partial\rho}{\partial t}+\bfnabla\cdot(\rho \bv)&=&0, \nonumber \\
\frac{\partial( \rho\bv)}{\partial t}+\bfnabla\cdot({\rho \bv\bv-\bb\bb+{{\sf P}^{\ast}}}) &=&-\Prat\bm{ S_r}(\bP),\  \nonumber \\
\frac{\partial{E}}{\partial t}+\bfnabla\cdot\left[(E+P^{\ast})\bv-\bb(\bb\cdot\bv)\right]&=&-\Prat\Crat S_r(E),  \nonumber \\
\frac{\partial\bb}{\partial t}-\bfnabla\times(\bv\times\bb)&=&0.
\label{MHDequation}
\end{eqnarray}
The dimensionless source terms are
\begin{eqnarray}
S_r(E)&=&\sigma_a\left(T^4-E_r\right)\nonumber\\
&+&(\sigma_a-\sigma_s)\frac{\bv}{\Crat}\cdot\left(\bF_r
-\frac{\bv E_r+\bv\cdot{\sf P_r}}{\Crat}\right), \nonumber\\
\bm{S_r}(\bP)&=&-\left(\sigma_s+\sigma_a\right)\left(
\bF_r-\frac{\bv E_r+\bv\cdot{\sf P_r}}{\Crat}\right) \nonumber\\
&+&\frac{\bv}{\Crat}\sigma_a\left(T^4-E_r\right).
\label{dimensionlesssource}
\end{eqnarray}
The dimensionless form of the equations will be used in the remainder of the paper.

Unlike the radiation moment equations, which require a closure (for example the VET),
this set of equations is closed. However, in order to compare with previous results
based 
on solution of the radiation moment equations, we can still calculate the Eddington
tensor as
\begin{eqnarray}
{\sf f}=\frac{{\sf K}}{J}.
\end{eqnarray}

\section{Angular Discretization}
\label{sec:angles}
The same angular discretization scheme for the specific intensity 
as adopted by \cite{Davisetal2012} is used here. This is based on 
the original algorithm developed by \cite{Brulsetal1999}, which itself 
is based on the quadrature principle described in \cite{Carlson1963}. 
We only introduce the necessary definitions here, which will be used in 
the rest of the paper.

Each angle $\bn$ is uniquely specified by its direction cosines 
with respect to the three axes $\bn=(\mu_x,\ \mu_y,\ \mu_z)$, 
where $\mu_x^2+\mu_y^2+\mu_z^2=1$. A quadrature weight $W$ 
is assigned to each angle $\bn$, which represents the solid angle 
extended by the ray along direction $\bn$. Therefore, equation (\ref{integrateangle})
for the angular quadrature can be calculated as
\begin{eqnarray}
J&=&\sum_lI_lW_l, \nonumber\\
 H_i&=&\sum_l \mu_{l,i} I_lW_l, \nonumber\\
  K_{ij}&=&\sum_l \mu_{l,i}\mu_{l,j} I_lW_l,
\end{eqnarray}
where $i,\ j$ and $l$ represent three directions and all the angles in each cell respectively.
The most important feature of this angular discretization scheme for our algorithm is that for an isotropic 
distribution of specific intensity, this scheme guarantees $K_{ii}/J=1/3$ for any number of angles. 
This is necessary to avoid significant 
discretization errors when we calculate the radiation source terms (equation \ref{dimensionlesssource}) from 
the RT equation (\ref{RTequation}).

\section{Numerical Algorithm}
\label{sec:algorithm}
The existence of the time dependent terms makes the RT equations become an
initial value problem, instead of 
a boundary value problem as in the case of the time-independent equations solved
in \cite{Davisetal2012}. Therefore, the 
short characteristic method adopted by \cite{Davisetal2012} cannot be used here and a new numerical algorithm 
is required. We first give an overview of the steps in the method, and then describe
each step in detail subsequently. 

\subsection{Basic Steps in the Algorithm}

\emph{Step 1.} Calculate time step $\Delta t$ based on cell size and speed of light 
with CFL number $0.4$.

\emph{Step 2.} Calculate the change of specific intensity along each direction $\bn_l$ 
due to divergence of transport flux along three directions $\Delta F_l$, according to Section \ref{sec:transport}. 
Depending on whether the scattering
opacity is larger or smaller than the absorption opacity in each cell, $\Delta F_l$ is added to the 
right hand side of the matrix either in \emph{Step 5} or \emph{Step 4}. This is necessary to keep 
the balance between the transport term and source terms.

\emph{Step 3.} Estimate flow velocity for half time step according to Section \ref{sec:estimatev}.

\emph{Step 4.} Update all the specific intensities with the absorption opacity related terms 
according to Section \ref{sec:absorption}. Calculate the corresponding radiation energy and momentum 
source terms.

\emph{Step 5.} Repeat \emph{Step 4} but for the scattering opacity related terms. 

\emph{Step 6.} Update gas quantities using the usual {\sc Athena} algorithm described in \cite{Stoneetal2008}.

\emph{Step 7.} Add the radiation energy and momentum source terms calculated in \emph{Step 4} and \emph{Step 5} 
to the gas quantities.

\subsection{The Transport step}
\label{sec:transport}
The left hand side of equation (\ref{RTequation}) represents the advection of specific intensity at the speed of light 
along the direction specified by $\bn$ when the source term vanishes. Based on this idea, 
\cite{StoneMihalas1992} applied upwind monotonic interpolation methods to solve the transport step in one dimension. 
In this way, each $I$ is advected across the grid using fluxes calculated at the cell interface, which guarantees conservation of radiation energy and momentum to 
roundoff error.  A similar technique can be used in multi-dimensions with appropriate extensions.

In the non-relativistic case, the direction of the specific intensity can be assumed to be constant during the transport step. Therefore, the transport part 
of equation (\ref{RTequation}) can be rewritten in conservative form
\begin{eqnarray}
 \frac{\partial I}{\partial t}+\Crat\bfnabla\cdot\left(\bn I\right)&=& \nonumber \\
  \frac{\partial I}{\partial t}+\Crat\frac{\partial}{\partial x}\left(\mu_x I\right)+\Crat\frac{\partial}{\partial y}\left(\mu_y I\right)+\Crat\frac{\partial}{\partial z}\left(\mu_z I\right)&=&0.
 \label{decomposeI}
\end{eqnarray}
This equation suggests that we can decompose the transport of $I$ along direction $\bn$ into three transport 
steps along each axis and each of them can be calculated  with similar method as proposed by \cite{StoneMihalas1992}. 

The original algorithm developed 
by \cite{StoneMihalas1992} needs to be modified for the velocity dependent source terms in optically thick regime 
to reduce numerical diffusion, especially for the case with pure scattering opacity. 
In the Eulerian frame, the transport term $\Crat\bfnabla\cdot\left(\bn I\right)$ includes both the diffusive part and advective part caused by 
the flow velocity. The diffusion and advection speed can be much smaller than the 
 speed of light. Therefore, if we always use the speed of light as the transport speed as originally adopted 
 by \cite{StoneMihalas1992}, the amount of numerical diffusion can significantly overwhelm the physical flux. 
 A similar issue also exists when the two radiation moment equations are solved in the mixed frame as 
 discussed in the Appendix of \cite{Jiangetal2013b}. 
 Therefore, it is necessary to separate the diffusion and advection parts so that we can treat them accurately. 
 
 The velocity dependent source term that needs special treatment is
 \begin{eqnarray}
 IV\equiv3\bn\cdot\bv J.
 \end{eqnarray}
 We split the transport term as
\begin{eqnarray}
\Crat\bn\cdot\bfnabla I=\Crat\bn\cdot\bfnabla\tilde{I}+\bn\cdot\bfnabla\left(IV\right),
\label{transporttwoterms}
\end{eqnarray}
where $\tilde{I}\equiv I-IV/\Crat$.
Mathematically, the equation is unchanged. However, the numerical treatments of the 
two terms are different to significantly reduce numerical diffusion. 
The transport velocity for the first term is chosen to be $\alpha\Crat$ \citep[][]{Jiangetal2013b} instead of $\Crat$, where 
\begin{eqnarray}
\alpha&=&\sqrt{\frac{1-\exp{(-\tau)}}{\tau}}, \nonumber\\
\tau&\equiv&\left[10\Delta l \times \left(\sigma_a+\sigma_s\right)\right]^2.
\end{eqnarray}  
Here $\Delta l$ is the cell size. When optical depth per cell is larger than $1$, $\alpha\sim 1/[10\Delta l (\sigma_a+\sigma_s)]$ 
and the transport speed is reduced by a factor of $\alpha$. 
When the cell is optically thin, $\alpha\sim 1$ and the transport speed is unchanged.  
As discussed in \cite{Jiangetal2013b}, the numerical factor $10$ in the above equation is chosen so that the numerical diffusion does not affect the 
physical solution over a wide range of optical depth per cell in our tests, while still keeping the algorithm stable. 

The speed of light is no longer the characteristic speed for the second term $\bn\cdot\bfnabla(IV)$ in 
equation (\ref{transporttwoterms}). This term represents the advection of specific intensity along direction 
$\bn$ due to motion of the flow. Therefore, the transport speed for this term is flow velocity $\bv$.

The two transport terms can both be decomposed along the three axes independently as 
\begin{eqnarray}
  \frac{\partial I}{\partial t}+\Crat\frac{\partial}{\partial x}\left(\mu_x\tilde{I}\right)+\Crat\frac{\partial}{\partial y}\left(\mu_y\tilde{I}\right)+\Crat\frac{\partial}{\partial z}\left(\mu_z\tilde{I}\right)=0. \nonumber \\
 \end{eqnarray}
 
 \begin{eqnarray}
 \frac{\partial I}{\partial t} +\frac{\partial}{\partial x}\left(\mu_x IV\right)+\frac{\partial}{\partial y}\left(\mu_y IV\right)+\frac{\partial}{\partial z}\left(\mu_z IV\right)&=& 0.
\end{eqnarray}
Although the above two equations are no longer the simple advection equation solved by \cite{StoneMihalas1992}, similar techniques can still be used to calculate 
the transport flux, as they represent the same advection process with advection velocity $\alpha\Crat$ and $|\bv|$ respectively, which are assumed to be 
constants during this step. The second order van Leer scheme as equation (4) of \cite{StoneMihalas1992} is adopted to calculate the flux.

\subsection{Absorption Opacity}
\label{sec:absorption}
The absorption opacity related terms change the specific intensity according to  
the following equation 
\begin{eqnarray}
 \frac{\partial I}{\partial t}&=&\Crat\sigma_a\left(\frac{T^4}{4\pi}-I\right)
 +3\bn\cdot\bv\sigma_a\frac{T^4}{4\pi}
 +\bn\cdot\bv\sigma_a I \nonumber \\
 &-&\sigma_a\frac{\bv\cdot\bv}{\Crat}J- \sigma_a\frac{\bv\cdot(\bv\cdot{\sf K})}{\Crat}.
 \label{absorptionequation}
\end{eqnarray}
The zeroth and first angular moments of the above equation multiplied by $4\pi$ give the following 
equations for the change of radiation energy density and flux as
\begin{eqnarray}
\frac{\partial E_r}{\partial t}=\Crat\sigma_a(T^4-E_r)+\sigma_a\bv\cdot\left(\bF_r
-\frac{\bv E_r+\bv\cdot{\sf P_r}}{\Crat}\right),\nonumber\\
\frac{\partial\bF_r}{\partial t}=-\Crat\sigma_a\left(
\bF_r-\frac{\bv E_r+\bv\cdot{\sf P_r}}{\Crat}\right)+\bv\sigma_a\left(T^4-E_r\right). \nonumber \\
\label{absorptionsource}
\end{eqnarray}
The corresponding changes of gas temperature and momentum are 
\begin{eqnarray}
\frac{\rho R_{\text{ideal}}}{\gamma-1}\frac{\partial T}{\partial t}&=&-\Prat\Crat\left(1-\frac{v^2}{\Crat^2}\right)\sigma_a\left(T^4-E_r\right) \nonumber \\
&-&2\Prat\sigma_a\bv\cdot\left(\bF_r
-\frac{\bv E_r+\bv\cdot{\sf P_r}}{\Crat}\right), \nonumber\\
\frac{\partial \rho\bv}{\partial t}&=&\Prat\sigma_a\left(\bF_r
-\frac{\bv E_r+\bv\cdot{\sf P_r}}{\Crat}\right)\nonumber \\
&-&\Prat\sigma_a\left(T^4-E_r\right)\frac{\bv}{\Crat}.
\end{eqnarray}
Notice that the change of kinetic energy density due to the work done by radiation force is
\begin{eqnarray}
\frac{\partial\left( \rho v^2/2\right)}{\partial t}&=&\Prat\sigma_a\bv\cdot\left(\bF_r
-\frac{\bv E_r+\bv\cdot{\sf P_r}}{\Crat}\right)\nonumber \\
&-&\Prat\sigma_a\left(T^4-E_r\right)\frac{v^2}{\Crat}.
\end{eqnarray}
Therefore, the change of total energy $\partial\left[\Prat E_r+\rho R_{\text{ideal}}T/(\gamma-1)+\rho v^2/2\right]/\partial t=0$ 
as the source terms we add to the gas and radiation have exactly the same value but opposite signs. 
As the thermalization time scale $\sim 1/\left(\Prat\Crat\sigma_a\right)$  can be 
very short even compared with light crossing time per cell 
when the optical depth per cell is much larger than $1$, equation (\ref{absorptionequation}) 
must be solved implicitly to make the algorithm stable. When the flow velocity is not too small compared with 
speed of light \footnote{For example, when $v/\Crat\sim 0.1$, the Lorentz factor is only 1.005 and 
relativistic effect is still negligible.} and radiation energy density is much larger than the gas pressure, 
the radiation work term is not negligible and should be solved 
simultaneously with other terms to get the correct equilibrium state. To avoid extra non-linear terms caused by the 
velocity dependent terms, flow velocity is fixed during this step (see section \ref{sec:estimatev}).
The absorption opacity is also fixed during this step. To get the correct equilibrium state, change of the gas temperature 
also needs to be calculated simultaneously with equation (\ref{absorptionequation}). Therefore, the specific intensity along 
each direction $\bn_m$ ($I^{n+1}_m$) and gas temperature ($T^{n+1}$) at time step $n+1$ can be calculated based on 
their values at current time step $n$ and time step $\Delta t$ as
\begin{eqnarray}
\frac{I_m^{n+1}-I_m^n}{\Delta t}&=&\Crat\sigma_a\left(\frac{(T^{n+1})^4}{4\pi}-I_m^{n+1}\right)\nonumber \\
&+&3\bn_m\cdot\bv\sigma_a\frac{(T^{n+1})^4}{4\pi}
+\bn_m\cdot\bv\sigma_a I_m^{n+1} \nonumber \\
&-&\sigma_a\frac{v^2}{\Crat}\sum_l\left(I^{n+1}_lW_l\right) \nonumber \\
&-&\frac{\sigma_a}{\Crat}\sum_l\left\{
\bv\cdot\left(\bv\cdot\bn_l\bn_l\right)I^{n+1}_lW_l\right\},\nonumber\\
\frac{\rho R_{\text{ideal}}}{\gamma-1}\frac{T^{n+1}-T^n}{\Delta t}&=&-8\pi\Prat\sigma_a\bv\cdot\left\{
\sum_l \left(\bn_l I_l^{n+1}W_l\right) \right . \nonumber \\
&-&\left .\frac{1}{\Crat}\sum_l \left(\bv I^{n+1}_l W_l  +\bv\cdot\bn_l\bn_l I^{n+1}_lW_l\right)\right\}\nonumber\\
&-&\Prat\Crat\left(1-\frac{v^2}{\Crat^2}\right)\sigma_a\left\{ (T^{n+1})^4 \right. \nonumber \\
&-&\left. 4\pi\sum_l\left(I_l^{n+1}W_l\right)\right\}.
\label{absorptionmatrix}
\end{eqnarray}
Here ranges for the lower scripts $m$ and $l$ are from $1$ to the total number of angles in each cell. 
For total $N$ angles for each cell, equation (\ref{absorptionmatrix}) represents
$N+1$ equations, which we have to solve simultaneously. 
This set of equations are non-linear because of the $(T^{n+1})^4$ term. They are linear with respect to $I^{n+1}_m$. We use Newton-Raphson iteration 
to solve this set of equations with the initial guess chosen to be $I^n_m$ and $T^n$. 
Because the non-linear terms are only fourth order polynomial, we usually find the iteration converges 
very quickly. 

Newton-Raphson iteration requires inversion of the Jacobi matrix during each step, which can be simplified analytically 
to significantly reduce the computational cost. We first subtract the line of $I^{n+1}_N$ from 
all the lines for $I^{n+1}_m (m\neq N)$. Each $I^{n+1}_m (m\neq N)$ can now be expressed as a function of $I^{n+1}_N$ and $T^{n+1}$ easily. 
Therefore we only need to invert a $2\times 2$ matrix for $I^{n+1}_N$ and $T^{n+1}$ numerically, after which all the $I^{n+1}_m (m\neq N)$ can be calculated 
directly. The total cost to invert the Jacobi matrix is only $\mathcal{O}(N)$.

With the updated solution $I^{n+1}_m$ and $T^{n+1}$, the changes of the gas internal energy density $\Delta E_T$ and gas momentum $\Delta(\rho\bv)$ 
due to the absorption opacity terms are
\begin{eqnarray}
\Delta E_T=\frac{\rho R_{\text{ideal}}}{\gamma}\left(T^{n+1}-T^n\right), 
\end{eqnarray}
\begin{eqnarray}
\Delta (\rho\bv)=-\frac{\Prat}{\Crat}\left[\sum_l\left(
\bn_l I_l^{n+1}W_l\right)-
\sum_l\left(\bn_l I_l^{n}W_l
\right)
\right].
\label{momentumchange}
\end{eqnarray}
The related change of kinetic energy density is
\begin{eqnarray}
\Delta E_{K}=\left((\rho\bv^n+\Delta (\rho\bv))^2-(\rho\bv^n)^2\right)/(2\rho).
\label{kinetichange}
\end{eqnarray}
We calculate the gas momentum source term $\Delta (\rho\bv)$ according to the conservation of gas 
and radiation momentum at time step $n$ and $n+1$, instead of calculating the momentum source 
term directly according to equation (\ref{absorptionsource}). This not only guarantees momentum conservation 
to roundoff error but also avoids difficulties when absorption opacity $\sigma_a$ is very large while $\bF_r$ is very small.
However, the sum of $\Delta E_T$ and $\Delta E_K$ differs from the change of radiation 
energy density on the order of $\mathcal{O}\left(\Delta t^2\right)$.  In principle, this energy error can be reduced when we go to second order accuracy in time, which requires solving equation (\ref{absorptionmatrix}) twice. 
Our approach to calculate $\Delta E_T$ and $\Delta E_{K}$ assures that gas temperature is evolved exactly as equation 
(\ref{absorptionmatrix}) describes and it will not be affected by the truncation error of the kinetic energy density. As we will show in the tests, the energy error is usually smaller than $0.01\%$ 
and it will not accumulate.

\subsection{Scattering Opacity}
\label{sec:scattering}
The scattering opacity does not change the gas temperature. It only changes the flow velocity and thus the kinetic energy density. The equations we need to solve for the 
scattering opacity related terms are
\begin{eqnarray}
 \frac{\partial I}{\partial t}&=&\Crat\sigma_s(J-I) +\bn\cdot\bv\sigma_s\left(I+3J \right)
 -2\sigma_s\bv\cdot\bH \nonumber \\
 &+&\sigma_s\frac{\bv\cdot\bv}{\Crat}J 
 +\sigma_s\frac{\bv\cdot(\bv\cdot{\sf K})}{\Crat}.
 \label{scatteringequation}
\end{eqnarray}
Taking the zeroth and first moments of above equation and multiply by $4\pi$, we get the equations describing the evolutions of radiation energy 
density and momentum as
\begin{eqnarray}
\frac{\partial E_r}{\partial t}&=&-\sigma_s\bv\cdot\left(\bF_r
-\frac{\bv E_r+\bv\cdot{\sf P_r}}{\Crat}\right), \nonumber\\
\frac{\partial\bF_r}{\partial t}&=&-\Crat\sigma_s\left(
\bF_r-\frac{\bv E_r+\bv\cdot{\sf P_r}}{\Crat}\right).
\end{eqnarray}
The change of gas momentum due to scattering opacity is
\begin{eqnarray}
 \frac{\partial \rho\bv}{\partial t}=\Prat\sigma_s\left(\bF_r
-\frac{\bv E_r+\bv\cdot{\sf P_r}}{\Crat}\right).
\end{eqnarray}
The velocity dependent terms in equation (\ref{scatteringequation}) also needs to be calculated 
simultaneously with the other term in equation (\ref{scatteringequation}) in order to get the correct 
equilibrium state, especially when the velocity dependent terms become significant. In order to make 
the code stable for the regime when the scattering optical depth per cell is larger than $1$, the 
scattering opacity source terms also need to be added implicitly. 

Following the approach used in section \ref{sec:absorption} for the absorption opacity case, we fix flow velocity during this 
step to avoid the nonlinear terms. Then each specific intensity $I_m$ is updated using backward Euler implicitly as
\begin{eqnarray}
\frac{I_m^{n+1}-I_m^n}{\Delta t}&=&\Crat\sigma_s\left(
\sum_l \left(I_l^{n+1}W_l\right)-I_m^{n+1}
\right)\nonumber \\
&+&\bn_m\cdot\bv\sigma_s\left(I_m^{n+1}+3\sum_l\left(I_l^{n+1}W_l\right)\right)\nonumber\\
&-&2\sigma_s\sum_l\left(\bv\cdot\bn_l I_l^{n+1}W_l\right)
+\sigma_s\frac{v^2}{\Crat}\sum_l\left(
I^{n+1}_lW_l\right)\nonumber\\
&+&\frac{\sigma_s}{\Crat}\sum_l\left(
\bv\cdot\left(\bv\cdot\bn_l\bn_l I^{n+1}_lW_l\right)\right).
\end{eqnarray}
These $N$ equations are linear with respect to the unknowns $I^{n+1}_m$ for given $I^{n}_m$, velocity and opacity. 
We find it is actually easier to use $I^{n+1}_mW_m$ as unknowns and the $N\times N$ matrix we need to invert has the following format
\begin{eqnarray}
{\sf M_s} = \left[\begin{smallmatrix} 
a_1+b_1+c_1&b_2+c_1&\cdots & \cdots & b_N+c_1 \\
b_1+c_2&a_2+b_2+c_2&\cdots & \cdots & b_N+c_2 \\
\vdots &  \vdots                 & \vdots & \vdots & \vdots  \\
b_1+c_N & b_2+c_N  &   \cdots  & \cdots & a_N+b_N+c_N 
\end{smallmatrix}
\right].
\end{eqnarray}
The matrix elements $a_m$, $b_m$ and $c_m$ are
\begin{eqnarray}
a_m&=&\frac{1}{W_m}\left(1+\Delta t\Crat\sigma_s-\Delta t\sigma_s\bn_m\cdot\bv\right),\nonumber\\
b_m&=&-\Delta t\Crat\sigma_s+2\Delta t\sigma_s\bv\cdot\bn_m \nonumber\\
&-&\Delta t \sigma_s\frac{v^2}{\Crat}-\Delta t\frac{\sigma_s}{\Crat}\left(\bv\cdot\left(\bv\cdot\bn_l\bn_l\right)\right),\nonumber\\
c_m&=&-3\Delta t\bn_m\cdot\bv\sigma_s.
\end{eqnarray}
LU decomposition is used to invert the matrix. Because of the special pattern of the matrix, the LU decomposition can be 
worked out analytically. The L and U matrix after the decomposition only have two lines that are non-zero. Therefore, the final cost 
to invert this matrix is still only $\mathcal{O}(N)$.
After we get the updated solution $I^{n+1}_m$, we calculate the change of gas momentum 
and kinetic energy density according to equations (\ref{momentumchange}) and (\ref{kinetichange}).

\subsection{Estimating the Velocity}
\label{sec:estimatev}
When source terms are added as described in section \ref{sec:absorption} and \ref{sec:scattering}, the flow velocity is fixed to avoid the highly 
non-linear terms during the matrix inversions. When photon momentum is significant compared with gas momentum, this is no longer 
a good approximation. To improve the accuracy, instead of using the flow velocity at time step $n$,  we first estimate the flow velocity 
at time step $n+1/2$ based on the following equations
\begin{eqnarray}
\frac{\partial \rho \bv}{\partial t}&=&\Prat\left(\sigma_s+\sigma_a\right)\left(
\bF_r-\frac{\bv E_r+\bv P_r}{\Crat}\right),\nonumber\\
\frac{\partial \bF_r}{\partial t}&=&-\Crat\left(\sigma_s+\sigma_a\right)\left(
\bF_r-\frac{\bv E_r+\bv P_r}{\Crat}\right),
\end{eqnarray}
where $P_r$ is the diagonal components of the radiation pressure tensor ${\sf P_r}$. We keep $E_r$ and $P_r$ fixed in this step and solve 
$\bv$ and $\bF_r$ from the above two equations. For given values of $\bv^{n}$ and $\bF_r^{n}$, we estimate the velocity $\tilde{\bv}$ according to
\begin{eqnarray}
\rho\tilde{\bv}-\rho\bv^{n}&=&0.5\Delta t\Prat\left(\sigma_s+\sigma_a\right)\left(
\frac{\Crat}{\Prat}\rho\bv^n +\bF_r^{n} \right.\nonumber \\
&-&\left.\frac{\Crat}{\Prat}\rho\tilde{\bv}-\frac{\tilde{\bv}}{\Crat}\left(E_r+P_r\right)
\right).
\end{eqnarray}
The estimated flow velocity $\tilde{\bv}$ is used when the absorption and scattering source terms are added.
 
 \section{Verification Tests}
 \label{sec:test}

\subsection{Thermal Equilibrium in A Uniform Static Medium}
\label{sec:static}

\begin{figure*}[htp]
\centering
\includegraphics[width=0.9\hsize]{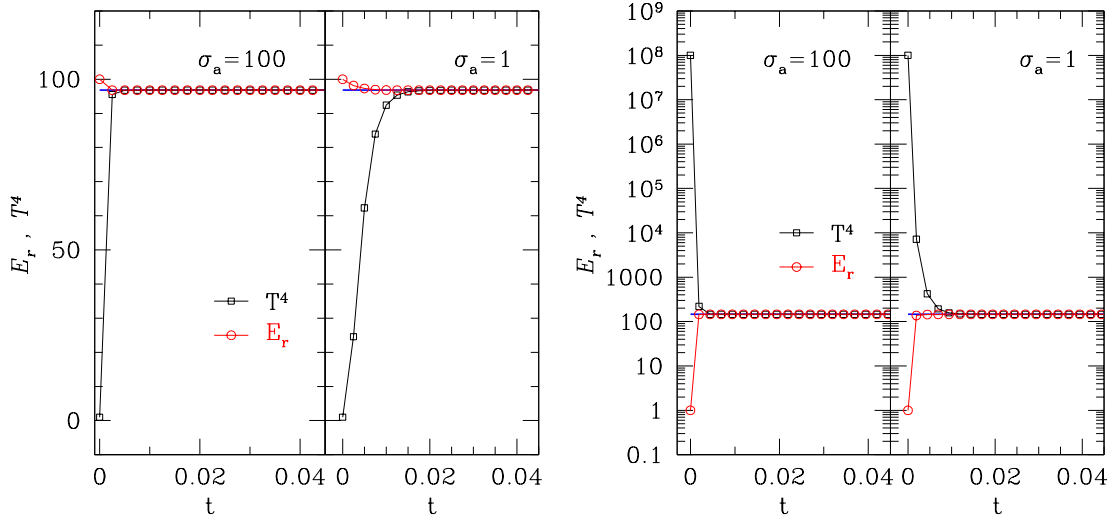}
\caption{Evolution of the radiation energy density and gas temperature to an equilibrium state in a static medium.
Different initial conditions are used for the left and right panels. For each initial condition, we show results with two different 
opacities as labeled in the figure. The red (black) lines are for $E_r$ ($T^4$) while the blue lines are the analytic solution at 
the equilibrium state. The black squares and red circles indicate the values at each time step.}
\label{thermalequilibrium}
\end{figure*}

\begin{figure}[htp]
\centering
\includegraphics[width=1.0\hsize]{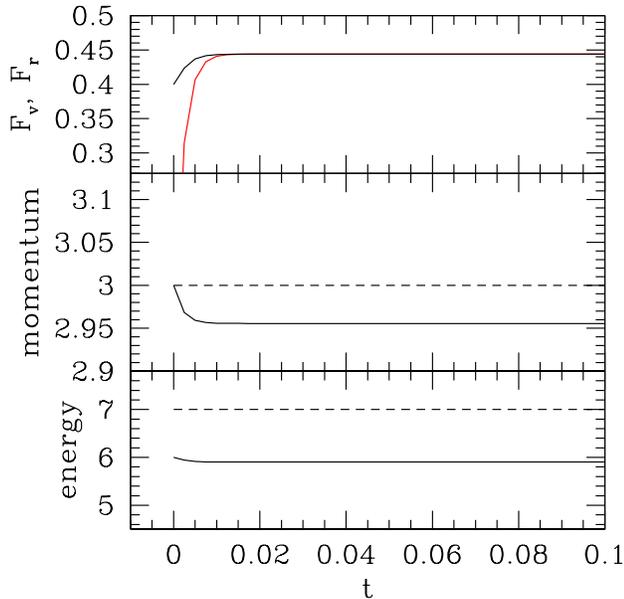}
\caption{{\emph Top:} History of the radiation flux $F_{r,x}$ (red line) and advective flux $F_v$ (black line) 
along $x$ direction for the test discussed 
in Section \ref{sec:beaming} for the case with absorption opacity. {\emph Middle:} History of the gas momentum (solid line) 
and total momentum (dashed line). {\emph Bottom:} History of the gas (solid line) 
and total energy densities (dashed line).}
\label{Radmomentum}
\end{figure}

\begin{figure*}[htp]
\centering
\includegraphics[width=1.0\hsize]{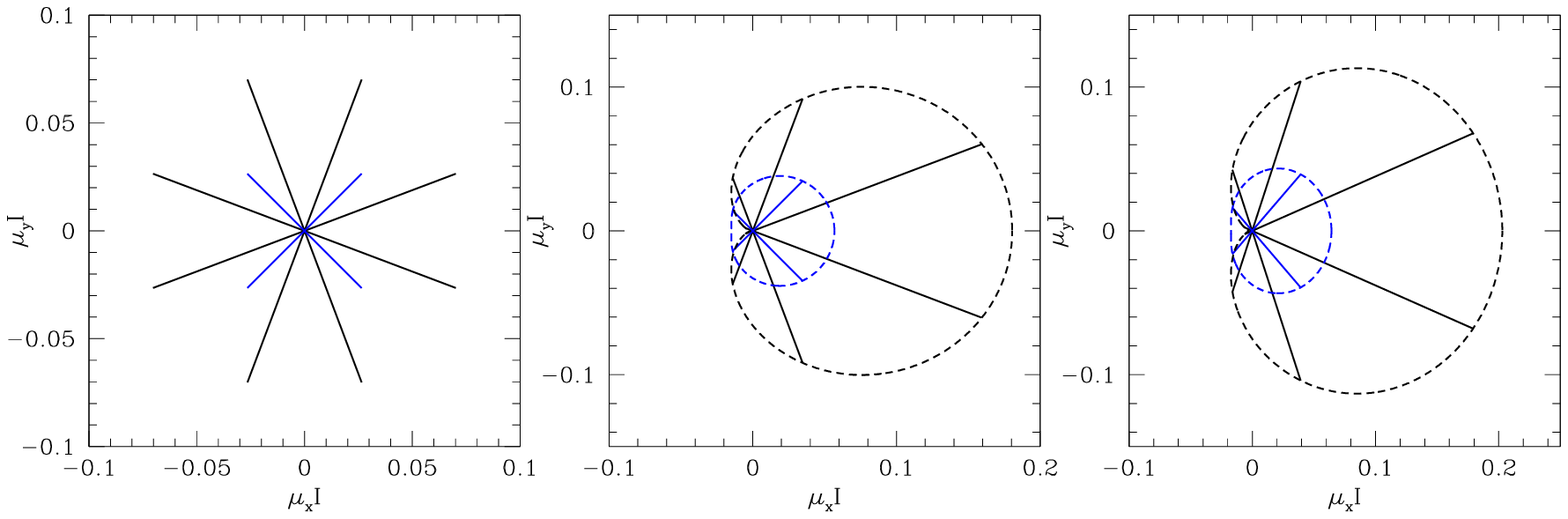}
\caption{Spatial distributions of the specific intensity projected in the $x-y$ plane for the tests described in Section \ref{sec:beaming}. 
The black rays have $\mu_z=0.333$ while for the blue rays $\mu_z=0.882$. 
\emph{Left:} Initial isotropic distribution. \emph{Middle:} Distribution of the specific intensities
at the equilibrium state with absorption opacity. \emph{Right:} The same as the 
middle panel but for the case with only scattering opacity. The black and blue ellipses are the analytical solutions for all the angles. }
\label{beamingmap}
\end{figure*}

To demonstrate that our treatment of absorption opacity as described in Section \ref{sec:absorption} can give 
the correct solution for radiation energy density $E_r$ and gas temperature $T$, we first study the 
evolution of $E_r$ and $T$ towards an equilibrium state in a static medium.  It is
important to show 
that even when the time step is much larger than the thermalization time, the method is 
still stable and accurate. 

We setup a uniform 2D domain, periodic in $X$ and $y$, with
$L_x=L_y=1$ and $N_x=N_y=32$. Density is chosen to be $1$ and flow
velocity is zero. The dimensionless parameters are $\Prat=1, \Crat=10,
R_{\text{ideal}}=1$ and $\gamma=5/3$.  We choose initial conditions
with $E_r \ne T^4$ and follow it as it relaxes to a state with $E_r =
T^4$, subject to energy conservation.  Figure \ref{thermalequilibrium}
shows initial conditions $E_r=100, T^4=1$ for the left panel and
$T^4=10^8, E_r=1$ for the right panel. Evolution of $E_r$ and $T$ for
the two different initial conditions with two different opacities
$\sigma_a$ are shown in Figure \ref{thermalequilibrium}. The
equilibrium solutions from the code agree with the analytical
solutions (the blue lines) very well. The total energy error in this
test is smaller than $10^{-10}$, which is set by the accuracy of the
Newton-Raphson iterations as discussed in Section
\ref{sec:absorption}.

When $\sigma_a=100$, the time step, which is set by the speed of light, is much
larger than the thermalization time. In this case, the correct equilibrium state
is achieved within about one time step. 
When the time step is comparable to the thermalization time as in the case $\sigma_a=1$, our algorithm automatically 
resolves the thermalization process. Unlike the predict and correct scheme shown in  Figure 1 and Figure 2 
of JSD12, this algorithm 
can always keep the relative values of $E_r$ and $T^4$ even for extreme parameters during the thermalization process, 
as our algorithm solves the gas temperature and radiation energy density simultaneously.

\subsection{Equilibrium State in A Moving Medium}
\label{sec:beaming}
To test our treatment of the velocity dependent terms, we give the fluid a uniform initial velocity $v_x$ along the $x$ direction 
based on the setup used in the last section and let the system evolve. 
As the RT equations are not Galilean invariant \citep[][JSD12]{Lowrieetal1999}, specific intensity will be 
beamed along the direction of flow velocity and a net radiation flux is produced along the opposite direction.
As the co-moving flux is non-zero initially, the flow experiences a drag by the photons and is decelerated until 
the co-moving flux becomes zero in the equilibrium state. We use $v_x=0.3\Crat$ in this test so that the 
velocity dependent terms are significant. For this test, we set $T=1$ and $E_r=1$ initially. Three angles for each octant are used. 

Histories of the flux, momentum and energy for the case with absorption opacity $\sigma_a=100$ are shown in Figure \ref{Radmomentum}. 
The advective flux is defined to be $F_v\equiv v_x\left(E_r+ P_{r,xx}\right)/\Crat$. When the system reaches the equilibrium state, 
$F_v=F_r$ and the co-moving flux becomes zero, as it should be. The total momentum of the system is the sum of the gas momentum 
$\rho v_x$ and photon momentum $\Prat F_r/\Crat$, which is conserved as shown in the middle panel of Figure \ref{Radmomentum}. 
The gas momentum is decreased due to the radiation drag and converted to photon momentum. During this process, the kinetic 
energy of the gas is transformed to the radiation energy density and gas internal energy through the thermal coupling. As shown in the 
bottom panel of Figure \ref{Radmomentum}, the total energy is also almost conserved. The relative energy error is only $10^{-4}$. 
For the case with scattering opacity $\sigma_s=100$, we find very similar behavior and accuracy. The only exception is that the gas temperature is unchanged 
and all the kinetic energy density is converted to the radiation energy density, as expected. Therefore, solutions from our transfer equation are 
consistent with the radiation moment equations. Notice that the last two terms in equation \ref{RTequation}, which we added in comparison 
to \cite{MihalasKlein1982}, are crucial to get the correct equilibrium state in a moving medium. 

Specific intensities not only allow calculation of the radiation energy density and flux, but also encode information about the 
angular distribution of the photons.  An initially isotropic distribution 
of the specific intensity is shown in the left panel of Figure \ref{beamingmap}. After the equilibrium state is reached, 
the middle and right panels of Figure \ref{beamingmap} show the spatial distributions of the 
specific intensities projected to the $x-y$ plane for the cases with absorption opacity $\sigma_a=100$ and scattering opacity $\sigma_s=100$ respectively. 
The right-going intensities are increased while the left-going intensities are decreased. In other words, the 
specific intensities are beamed as expected. 

To check that the angular distribution of specific intensities agree with the equations (\ref{absorptionequation}) and 
(\ref{scatteringequation}) quantitatively in the steady state, we set the left hand side of the two equations to be zero and solve the two 
integral equations of the right hand sides for all angles, which are shown as the black and blue ellipses in Figure \ref{beamingmap}. 
As this figure shows, the agreement is perfect.

\begin{figure}[htp]
\centering
\includegraphics[width=1.0\hsize]{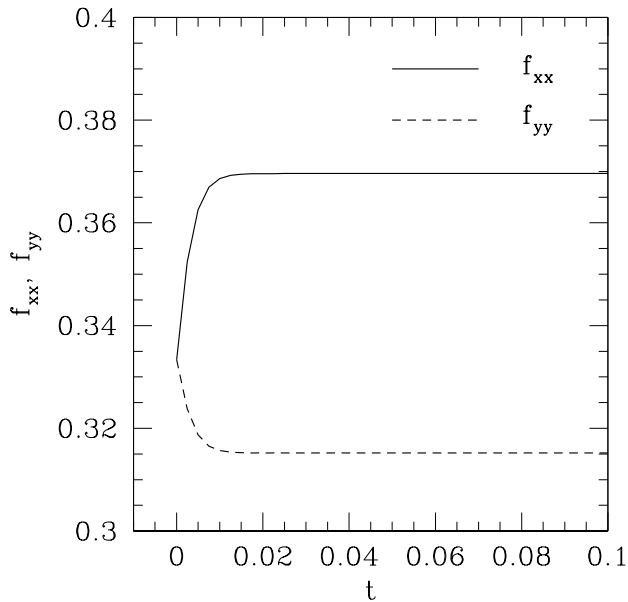}
\caption{Histories of the $x-x$ and $y-y$ components of the Eddington tensor for the test described in 
section \ref{sec:beaming} with pure absorption opacity. }
\label{Eddtensor}
\end{figure}

The angular distribution of the photons can also be quantified by two components of the Eddington tensor 
$f_{xx}\equiv P_{r,xx}/E_r$ and $f_{yy}\equiv P_{r,yy}/E_r$ in this test. Figure \ref{Eddtensor} shows the 
evolution of $f_{xx}$ and $f_{yy}$ for the case with pure absorption opacity. They are $1/3$ initially when  
the specific intensities are isotropic. In the equilibrium state, $f_{xx}$ is increased to $\sim 0.37$ 
while $f_{yy}$ drops to $\sim 0.315$, although the gas temperature is spatially isotropic. Consistent with Figure \ref{beamingmap}, 
the difference between $f_{xx}$ and $f_{yy}$ in the Eulerian frame is caused by the beaming of the fluid velocity. The 
degree of anisotropy is $\sim \mathcal{O}\left(v/\Crat\right)^2 $ in this thermal equilibrium case \citep[][]{MihalasMihalas1984}.

\subsection{Non-LTE Atmosphere}
\label{sec:atmosphere}

\begin{figure}[htp]
\centering
\includegraphics[width=1.0\hsize]{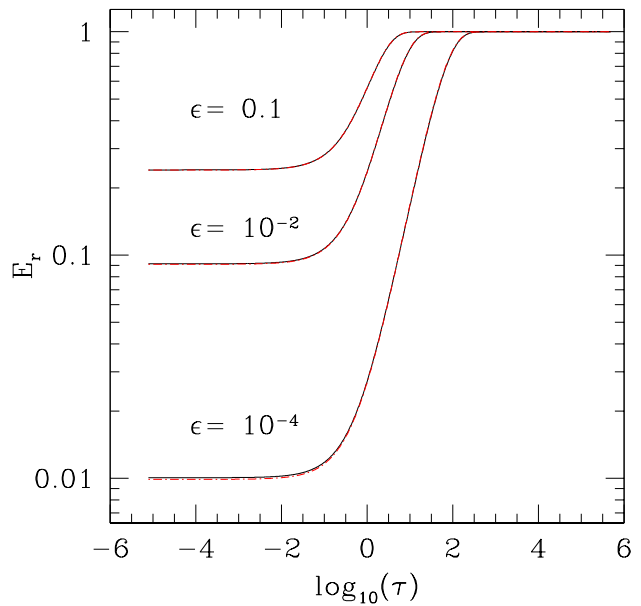}
\caption{Comparison of the profiles of the radiation energy density between the numerical (black solid lines) 
and analytical solutions (red dashed lines) for the non-LTE atmosphere test 
described in Section \ref{sec:atmosphere}. The value of $\epsilon$ is labeled above each line. } 
\label{atmosphere}
\end{figure}

The steady state solution in a scattering opacity dominated atmosphere is used to test the 
short characteristic module described in \cite{Davisetal2012}, which solves the time-independent 
transfer equation and finds the solutions via iterations. This test is useful as it covers the transition from optically thick LTE to 
optically thin non-LTE regimes within the atmosphere. 
It also demonstrates that steady state solution can be achieved by solving the \emph{time dependent} RT equations.

The simulation domain is a cubic box with $x\times y\times z\in \left(-0.5, 0.5\right)\times \left(-0.5, 0.5\right)\times\left(-10, 10\right)$. 
The spatial resolution is fixed to be $16\times 16\times 1280$ for the $x$, $y$ and $z$ directions respectively. 
Density of the atmosphere is chosen to be an exponential profile  along vertical direction $\rho(x,y,z)=10^{-3}\exp(-z+10)$. 
Gas temperature is fixed to be $1$ everywhere and flow velocity is zero everywhere. Absorption opacity is $\sigma_a=\epsilon \rho$ 
while the scattering opacity is $\sigma_s=(1-\epsilon)\rho$. The parameter $\epsilon$ is the destruction parameter used in 
\cite{Davisetal2012}, which is a constant here. If the Eddington tensor is fixed to be $1/3{\sf I}$, then the analytic 
solution for the radiation energy density is (equation 30 of \cite{Davisetal2012})
\begin{eqnarray}
E_r(\tau)=1-\frac{e ^{-\sqrt{3\epsilon}\tau}}{1+\sqrt{\epsilon}},
\end{eqnarray}
where $\tau$ is the optical depth measured from the top $z=10$. In order to be close to the Eddington approximation, we use one angle 
per octant. Periodic boundary conditions are used along the $x$ and $y$ directions. At the bottom of the simulation box, we copy the specific 
intensities along all angles from the last active zones to the ghost zones. At the top, only outgoing specific intensities are copied from 
the last active zones to the ghost zones while the incoming specific intensity is set to be zero in the ghost zones. The initial specific intensities 
are set to be $1/(4\pi)$ for all angles everywhere. The hydrodynamical equations
are not evolved in this test.
Because of the vacuum boundary condition at the top, the radiation energy density 
will decrease until cooling is balanced by the thermal emission in the equilibrium state. The smaller is $\epsilon$, the smaller is 
$E_r$ at the top. The solutions from our code at the equilibrium state for three different values of $\epsilon$ are shown as the solid 
black lines in Figure \ref{atmosphere}. Compared with the analytical solutions shown as the dashed red lines, they agree very well when 
$\epsilon$ changes from $0.1$ to $10^{-4}$.

\subsection{Crossing Beams in Vacuum}
\label{sec:twostreams}

\begin{figure}[htp]
\centering
\includegraphics[width=0.8\hsize]{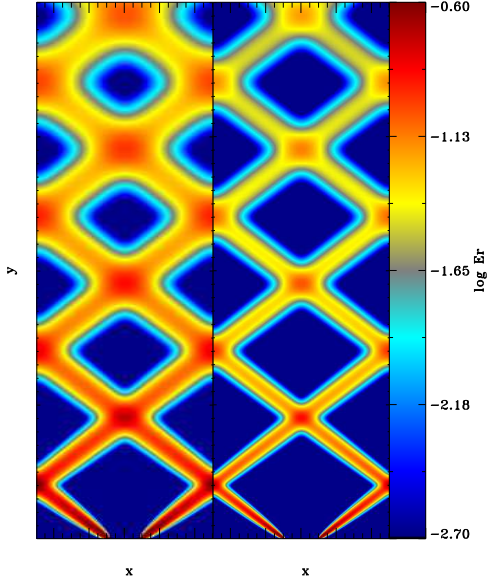}
\caption{Distribution of radiation energy density for two beams of photons crossing vacuum, as described in 
Section \ref{sec:twostreams}. The left panel is the result for a resolution of $N_x=128, N_y=512$ while this resolution is 
doubled for both directions in the right panel.  }
\label{twostreams}
\end{figure}

Unlike the short characteristic module used in \cite{Davisetal2012}, specific intensities are decomposed along each 
axis for the transport step. By propagating two beams of photons in vacuum, we demonstrate that this approach 
still allows accurate propagation of photons in a direction oblique to the grid. 

We setup a 2D domain with box size $L_x=1, L_y=4$. Periodic boundary 
conditions are used along the $x$ direction. Both absorption and scattering opacity are zero everywhere. We inject two beams 
from the bottom at $45$ degrees with respect to the $+x$ and $-x$ axes respectively. Because of the periodic boundary conditions, 
the two beams will cross each other and finally escape from the top. The distributions of $E_r$ after the two beams have escaped 
for two different spatial resolutions are shown in Figure \ref{twostreams}. Compared with Figure 6 of \cite{Davisetal2012}, our approach gives similar results to the 
short characteristic method with quadratic interpolation.  There is a small amount of
numerical diffusion in the beam, although total flux remains conserved.  The numerical
diffusion is reduced with increasing resolution, which reflects convergence of the
solution.

As an aside, we note that M1 methods fail this test as the two beams of radiation merge
into one at the point where they cross.

\subsection{Dynamic Diffusion}
\label{sec:advectiondiff}

\begin{figure}[htp]
\centering
\includegraphics[width=0.9\hsize]{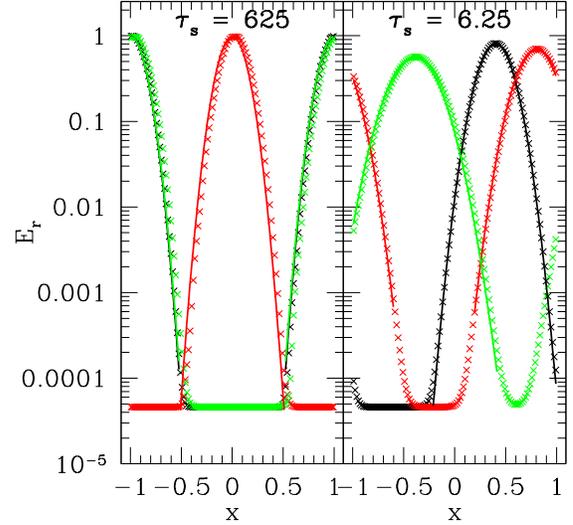}
\caption{Profiles of radiation energy density $E_r$ 
for the dynamic diffusion tests described in Section \ref{sec:advectiondiff}. 
\emph{Left:} Results when optical depth per cell is $\tau_s=625$. The black, red 
and green points are simulation results at times $t=1,2,3$ respectively.  \emph{Right:} Results when 
optical depth per cell is reduced to $\tau_s=6.25$. The time for the black, red and green dots are
$0.4, 0.8, 1.6$ respectively. The corresponding 
lines in the two panels are analytical solutions given by equation \ref{diffsolution}.
 }
\label{AdvectionDiff}
\end{figure}

\begin{figure}[htp]
\centering
\includegraphics[width=0.9\hsize]{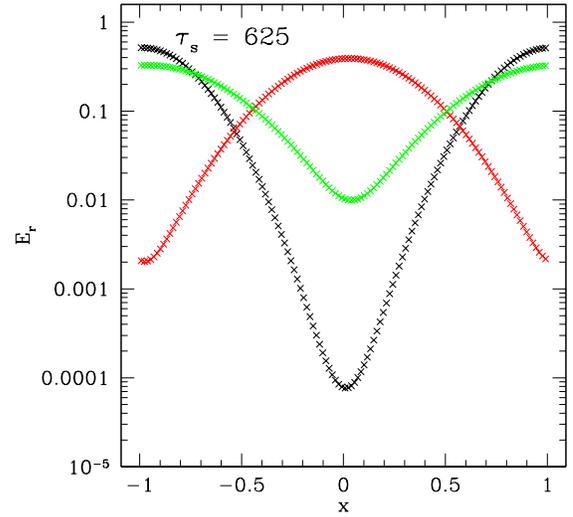}
\caption{Same as the left panel of Figure \ref{AdvectionDiff} but for the case 
when the original transport scheme given by \cite{StoneMihalas1992} 
is used.}
\label{AdvectionDiffWrong}
\end{figure}

As described in the Appendix of \cite{Jiangetal2013b}, the most useful test for our treatment of the transport step described 
in Section \ref{sec:transport} is the dynamic diffusion test in an optically thick medium with pure scattering opacity. We setup a 1D domain 
with $L_x=2$. The density, temperature and flow velocity are all set to be $1$. All the hydro 
quantities are uniform over the whole simulation domain and they do not evolve in the test. The dimensionless speed of 
light is $\Crat=10$. The value of $\Prat$ does not affect the test. The scattering opacity $\sigma_s$ is a parameter that controls 
the diffusion coefficient $D\equiv \Crat/(3\sigma_s)$. We use $128$ grid points
and periodic boundary 
conditions.

The initial profiles of radiation energy density and flux for the region $x\in (-0.5,0.5)$ are
\begin{eqnarray}
E_r(x,0)&=&\exp(-40 x^2),\nonumber\\
F_r(x,0)&=&\frac{80 x}{3\sigma_s}E_r(x,0)+\frac{4v}{3\Crat} E_r(x,0).
\end{eqnarray}
For other positions, $E_r$ and $F_r$ are fixed to be $\exp(-10)$ and 
$4vE_r/(3\Crat)$. We will only focus on the evolution of the region $x\in(-0.5+\sqrt{Dt}, 0.5-\sqrt{Dt})$, 
as this part will not be affected by the finite domain of the simulation box. In order to 
be consistent with the Eddington approximation, so that analytical solutions can be
used for comparison, we only use one angle per octant. 
Therefore the specific intensities are uniquely determined by $E_r$ and $F_r$. 

In the diffusion limit in the Eddington approximation, the profile of $E_r$ at time $t$ should be \citep[][]{SekoraStone2010, Jiangetal2013b}
\begin{eqnarray}
E_r(x,t)=\frac{1}{(160Dt+1)^{1/2}}\exp\left(\frac{-40(x-vt)^2}{160Dt+1}\right).
\label{diffsolution}
\end{eqnarray}
Comparisons between the numerical and analytical solutions at three different times for two different values of opacity $\sigma_s$ are 
shown in Figure \ref{AdvectionDiff}. This figure shows that our algorithm can calculate the diffusion and advection processes accurately 
for a wide range of opacity. In order to show the necessity of separating the
diffusive and advective terms as described in Section 
\ref{sec:transport}, we repeat the test with optical depth per cell $\tau_s=625$ 
and use the original transport scheme described by \cite{StoneMihalas1992}.
The result is shown 
in Figure \ref{AdvectionDiffWrong}. Compared with the left panel of Figure \ref{AdvectionDiff}, the numerical diffusion rate is clearly 
dominant over the physical diffusion rate and our treatment shows significant improvement.

\subsection{Linear Wave Convergence Test}
\label{sec:linearwave}

\begin{figure}[htp]
\centering
\includegraphics[width=0.9\hsize]{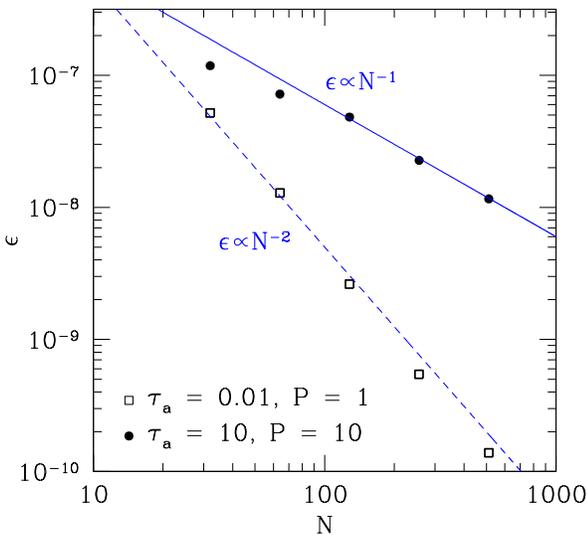}
\caption{Convergence of L1 error with resolution in the linear wave tests for two different set of 
parameters as labeled in the Figure. The solid and dashed blue lines are indications of first order 
and second order convergence for the optically thick and optically thin cases respectively. }
\label{Convergence}
\end{figure}

\begin{figure}[htp]
\centering
\includegraphics[width=0.9\hsize]{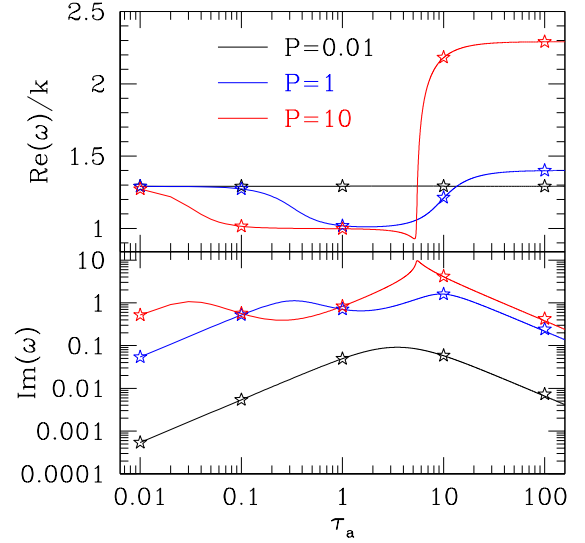}
\caption{Comparison of the numerically calculated phase velocity (top panel) and damping rate (lower panel) of linear acoustic waves 
with the analytical solution for a wide range of optical depth per wavelength and three different values of $\Prat$. 
The phase velocity is scaled to the isothermal sound speed and $\Crat=10$ for all the tests.
The stars represent quantities measured from simulations as described in Section \ref{sec:linearwave} while the lines are analytical 
solutions based on the dispersion relation given by \cite{Jiangetal2012}. The code gets the phase velocity and damping rate accurately 
for all the parameters we have explored.}
\label{Linearwave}
\end{figure}

In order to test the full radiation hydrodynamic algorithm, we carry out the linear wave test in a 2D domain based on the dispersion relation 
given by \cite{Jiangetal2012}. We setup a uniform medium with background state $\rho=T=P=1$ and $\gamma=5/3$. Radiation energy 
density $E_r=1$ and radiation flux is zero in the background state. In order to be consistent with the Eddington approximation used 
to calculate the dispersion relation, we use one angle per octant. The specific intensity in the background state is $I=1/(4\pi)$. We fix 
the dimensionless speed of light to be $\Crat=10$. The box size is $L_x$ and $L_y$ and the wave vector is aligned with $x$ axis in 
order to calculate the propagation speed and damping rate more easily.
As the hydrodynamics algorithm is second-order accurate, while the RT module is
only 
first-order accurate due to operator splitting, we expect that the overall
convergence rate for the whole code will depend on the parameter regime.  For
gas pressure dominated or optically thin regimes it
will be second-order accurate, while for optically thick 
radiation pressure dominated regimes it 
will be first-order accurate.
This is confirmed in Figure \ref{Convergence}. We show the change of L1 error  
with spatial resolution for two sets of parameters $\tau_a=0.01, \Prat=1$ and $\tau_a=10, \Prat=10$. When the optical depth per wavelength is 
only $0.01$, the code is second order accurate. When the optical depth per wavelength is increased to $10$ and radiation pressure is larger 
than gas pressure, the code is first order accurate. 

In Figure \ref{Linearwave}, we compare the wave propagation speed and damping rate from our code with the analytical solutions for 
optical depth per wavelength from $0.01$ to $100$ and ratio between radiation pressure and gas pressure from $0.01$ to $100$. 
In order to calculate the phase velocity and damping rate from our code, we evolve the linear wave for one period and calculate the 
Fourier transform of the density profile at the end of the simulation with background state subtracted. This is compared with the Fourier 
transform of the initial density profile with background state subtracted. We identify the positions of the component with maximum power 
in the phase space to be $\phi_0$ and $\phi_t$ for the initial and final solutions. The values of maximum power are $A_0$ and $A_t$ respectively. 
Therefore, the difference between the phase velocity from the code and the analytical solution is $\delta v=(\phi_t-\phi_0)/(kT)$, where $k=1/(2\pi)$ 
is the wave number and $T$ is the period. The damping rate in the code can be calculated as $\ln(A_0/A_t)/T$. The numerical values shown 
in Figure \ref{Linearwave} are calculated with $512$ grid points per wavelength. With this spatial resolution, our code can get the phase velocity and 
damping rate accurately over all the parameters we have explored. As $\Crat$ is only $10$, which is the regime that the algorithm will be 
most useful, we can easily be in the regime where photons are well-coupled to the gas,
and radiation pressure contributes a significant fraction of the 
restoring force for radiative acoustic modes. This is the case when $\tau_a>10$ with $\Prat=1$ and $\Prat=10$ in Figure \ref{Linearwave}, 
as the phase velocity is larger than the adiabatic sound speed for these cases. Again, our code captures these modes accurately.

\begin{figure}[htp]
\centering
\includegraphics[width=0.9\hsize]{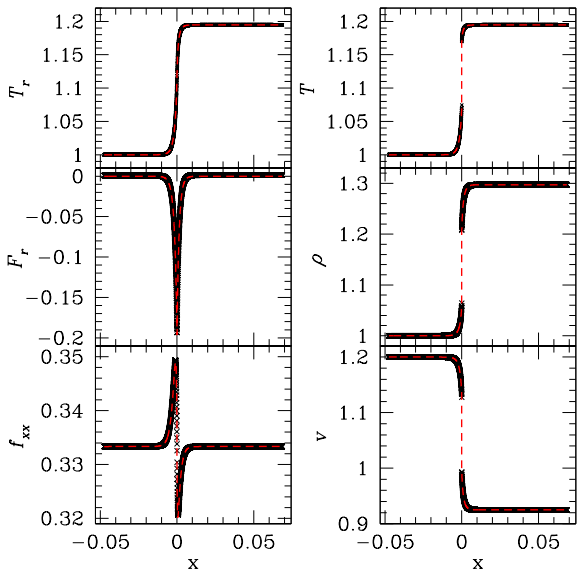}
\caption{Structure of a radiation-modified shock for Mach number 
$\mathcal{M}=1.2$. The dashed red lines are the semi-analytical solution by solving 
the time-independent radiation hydrodynamic equations while the black dots are the numerical 
results when the flow reaches steady state. }
\label{Shock1p2}
\end{figure}

\begin{figure}[htp]
\centering
\includegraphics[width=0.9\hsize]{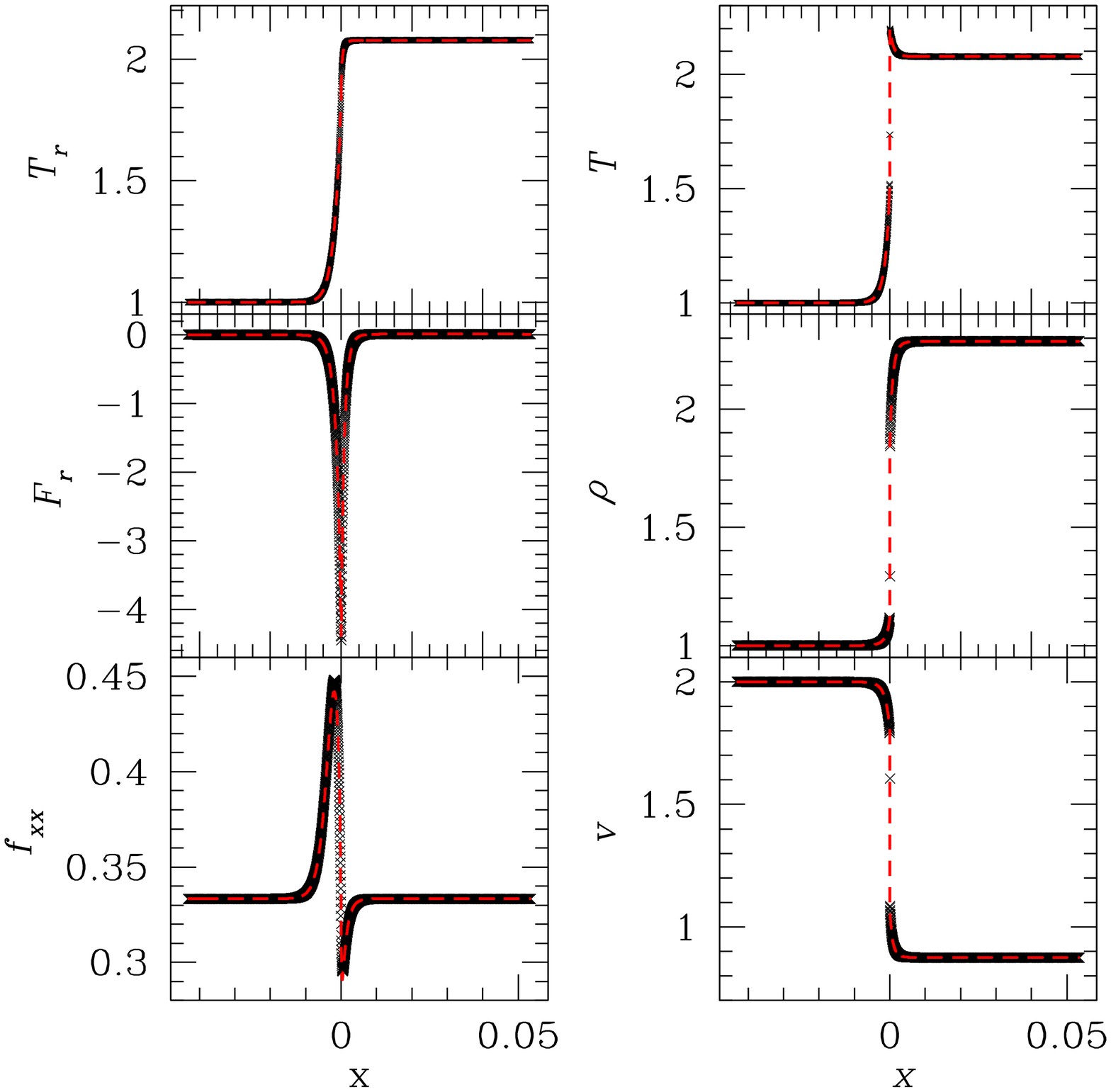}
\caption{The same as Figure \ref{Shock1p2} but for Mach number 
$\mathcal{M}=2$. }
\label{Shock2}
\end{figure}

\begin{figure}[htp]
\centering
\includegraphics[width=0.9\hsize]{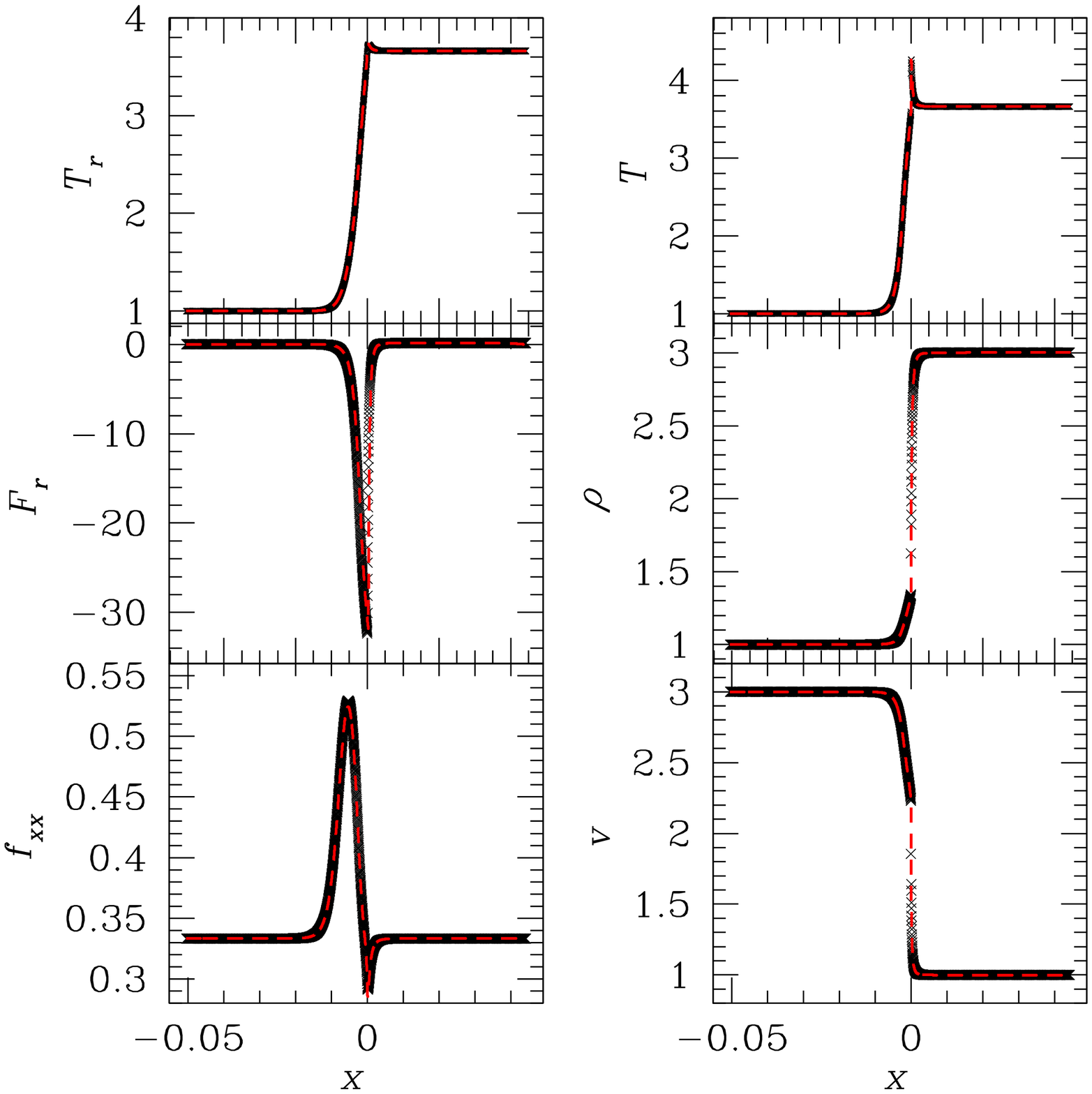}
\caption{The same as Figure \ref{Shock1p2} but for Mach number 
$\mathcal{M}=3$.  }
\label{Shock3}
\end{figure}

\begin{figure}[htp]
\centering
\includegraphics[width=0.9\hsize]{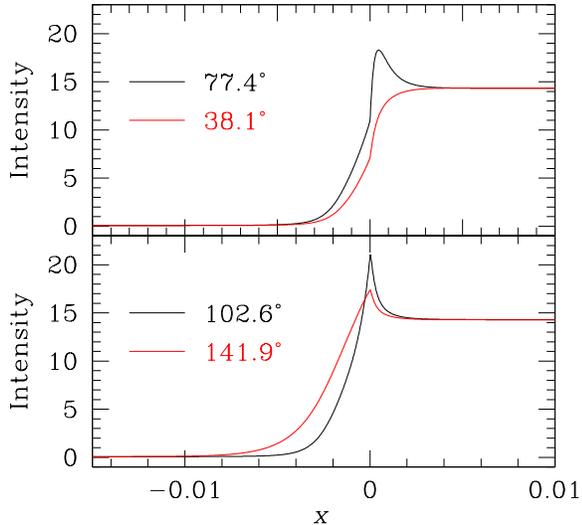}
\caption{Spatial profiles of the specific intensities at four different angles
for a radiation modified shock 
with Mach number $\mathcal{M}=3$. The angles labeled in each panel are with 
respect to $+x$ axis.  Note the radiation field is highly anisotropic. }
\label{Shock3Intensity}
\end{figure}

\subsection{Radiation Shock Test}
The effects of radiation on the structure of strong shocks have been used as standard 
tests for many radiation hydrodynamic codes \citep[][]{,Gonzalezetal2007,SekoraStone2010,Zhangetal2011,Jiangetal2012}. 
However, because of the complexity of the shock solutions, the numerical solutions are either compared 
with simple analytical estimates \citep[][]{Vaytetetal2013}, or semi-analytical 
solutions based on the Eddington approximation in more quantitative tests \citep[][]{LowrieEdwards2008, Jiangetal2012}.  
Recently, \cite{McClarrenDrake2010} have pointed out that the radiation field near the Zel'dovich spike in subcritical shocks can be 
very anisotropic, which invalidates the use of the Eddington approximation, and
may provides a good diagnostic to test the accuracy of more advanced RT
algorithms. 

Recently, the steady-state structure of radiation modified shocks at a variety
of Mach numbers has been worked
out without any assumption regarding the Eddington tensor 
by \cite{Ferguson2014}. We adopt these new solutions to test
our full RT and hydrodynamics algorithms\footnote{We thank 
Jim Ferguson for providing us his radiation shock solutions.} .

We initialize the shock solutions in 1D with dimensionless 
pre-shock parameters $\rho_0=T_0=E_{r,0}=1$ and $v_0=\mathcal{M}$, where the Mach number $\mathcal{M}$ 
controls the relative importance of radiation to gas pressure. 
The other dimensionless numbers are $\Crat=1.73\times10^3$, $\Prat=10^{-4}$, and
$\sigma_a=577.4$.  These values are
the same as in Section 6.2 of JSD12, so that the radiation shock solutions can be directly compared. Gas temperature 
$T$ is calculated as $T=P/(0.6\rho)$ and $\gamma=5/3$. We use $4096$ grid points
and $10$ angles per octant for all the solutions. 
All quantities on the left boundary are fixed to the pre-shock parameters, while on the right boundary all quantities in the 
ghost zones are copied from the last active zone. The steady state numerical solutions can be compared with the 
initial conditions to see how well the code can hold the semi-analytical solutions. 
Our numerical solutions after three flow crossing times for $\mathcal{M}=1.2,\ 2,\ 3$ are shown as 
black dots in Figures \ref{Shock1p2}, \ref{Shock2}, \ref{Shock3} while the semi-analytical solutions are shown as dashed red lines
in each figure. The numerical solutions agree with the semi-analytical solutions very well for the three cases. 

 In the case with $\mathcal{M}=1.2$, gas temperature increases monotonically from upstream to downstream. 
The $f_{xx}$ component of the Eddington tensor is close to $1/3$ near the shock front,
although it can be seen that 
$f_{xx}$ can be smaller than $1/3$ in the immediate post-shock flow, as found by \cite{Sincelletal1999} and JSD12.  
When the Mach number is increased to $2$ 
and then $3$, radiation pressure downstream becomes significantly stronger and the Zel'dovich spike \citep[][]{ZeldovichRaizer1967} emerges.
In Figure \ref{Shock3} with $\mathcal{M}=3$, not only does the gas temperature $T$ form a spike near the shock front, 
but also the radiation temperature $T_r$. At the same time, the radiation field becomes very anisotropic as the peak of 
$f_{xx}$ differs from $1/3$ by $50\%$. A similar spike in radiation temperature
is evident in Figure 15 
of JSD12 when the angular distributions of photons are calculated self-consistently with the VET approach. This feature is called 
``anti-diffusion" in \cite{McClarrenDrake2010}, because the direction of radiation flux at the shock front 
is along the same direction as the radiation energy density gradient.  This is the 
opposite behavior to what is assumed in the 
diffusion approximation.  It occurs because the RT equation requires $\bF_r\propto-\bfnabla\cdot{\sf P_r}=-\bfnabla\cdot\left({\sf f}E_r\right)$ 
in steady state. When the Eddington tensor has a strong spatial gradient as in the shock front, $\bF_r$ is not necessarily in the 
same direction of $-\bfnabla E_r$, as assumed in the diffusion approximation.
These results point out that it is crucial to adopt algorithms that can treat the
anisotropic nature of the radiation field accurately.

The anisotropic radiation field in our solution can also be demonstrated directly
by plotting
the specific intensities along different directions.
Profiles of specific intensity along four different angles for the case with 
Mach number $\mathcal{M}=3$ are shown in Figure \ref{Shock3Intensity}. The intensity
along rays pointing in the $-x$ direction is much 
larger than intensities in other directions in the upstream region. This is 
responsible for the pre-heating of the gas. The intensity along rays 
in the direction of the flow actually increase monotonically from upstream to downstream,
which is quite different from the profile 
of $E_r$ shown in Figure \ref{Shock3}. The spike in radiation energy density near the shock front 
is dominated by the rays perpendicular to the flow direction. This is because just in front of the hot downstream gas, rays 
perpendicular to the flow direction have a longer path length through the narrow shell
of hot gas, and therefore have proportionally more emitted photons than directions
with shorter path lengths through the emission region.
This is consistent with the fact that $f_{xx}<1/3$ in Figure \ref{Shock3} near the 
shock front.

\begin{figure}[htp]
\centering
\includegraphics[width=0.9\hsize]{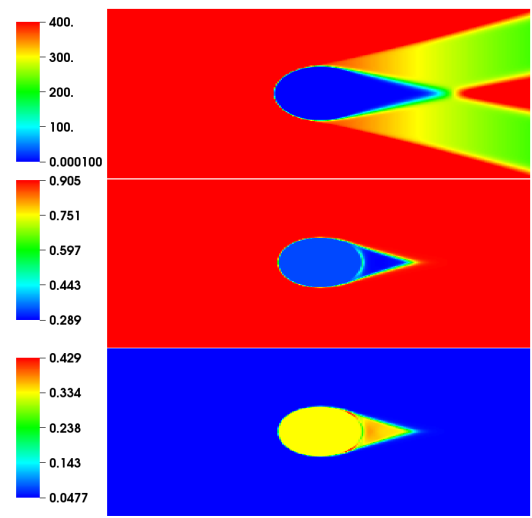}
\caption{Shadows created by an optically thick cloud from radiation beamed in two
directions
$\pm14^{\circ}$ with respect to the horizontal axis. From top to bottom, the three 
panels show radiation energy density $E_r$, $xx$ and $yy$ components of Eddington tensor. 
The umbra and penumbra behind the cloud are clearly visible. 
}
\label{Shadow}
\end{figure}

\subsection{Shadow Test}
The ability to capture shadows cast by an optically thick object in an optically thin environment 
requires propagating photons along different directions correctly. This test has been widely used to demonstrate the differences between 
FLD, M1 and RT algorithms based on short characteristics \citep[][JSD12]{HayesNorman2003,Gonzalezetal2007,McKinneyetal2013}. Although the M1 method is able to 
capture the shadow formed by one beam, therefore demonstrating an advantage of M1 over FLD \citep[][]{Gonzalezetal2007}, 
it cannot propagate two beams 
correctly \citep[][]{McKinneyetal2013}.  As our method directly solves for
the specific intensity along different directions, it should be 
able to capture the shadow from multiple beams accurately, as was the case with
the VET method in JSD12. 

For this test, we setup a rectangular box of size $(-0.5,0.5)\times(-0.3,0.3)$ with
dimensionless parameter $\Crat$ chosen to be $10$ to speed up the calculation. The background medium has 
density $\rho_0=1$ and temperature $T_0=1$. Density inside the elliptical 
region $r=x^2/a^2+y^2/b^2\leqslant 1$ with $a=0.1$ and $b=0.06$ is $\rho(r)=\rho_0+(\rho_1-\rho_0)/\left[1+\exp\left(10(r-1)\right)\right]$, 
where $\rho_1=10\rho_0$. Gas temperature is set to be $1/\rho$. Absorption opacity $\sigma_a=T^{-3.5}\rho^2$ is used in this test. 
Beamed photons with intensity $I=103.13$ are injected from the left $x$ boundary at angles $\pm14^{\circ}$ with respect to the $x-$axis. 
One beam at $-14^{\circ}$ and another beam at $14^{\circ}$ with the same
intensity are injected along the top and
bottom $y$ boundaries respectively.
The intensities in the ghost zones along the right $x$ boundaries are copied from the last active zones. We use a resolution of 
$512\times 256$ cells for this test. The hydrodynamical equations are not evolved 
(although see \cite{Progaetal2014} for the hydrodynamical evolution of an irradiated cloud computed using
our VET method).
Although we use different values of $\Crat$ compared with the shadow test shown in Section 6.5 of JSD12, all the other important 
dimensionless numbers and angles of the incoming intensities 
are the same. For example, the photon mean free path inside the cloud is only $3.2\times10^{-6}$ of the horizontal 
size of the simulation box due to the low temperature and high density, while it the same as the box size outside the cloud. 
Therefore, results from the two different RT algorithms can be directly compared. Radiation energy density, $xx$ and $yy$ components 
of the Eddington tensor after the photons 
have propagated through the simulation box are shown in Figure \ref{Shadow}. Compared with Figure 19 of JSD12, the umbra 
and penumbra calculated by the two codes are indeed very similar.
Note also the results differ significantly from those shown by \cite{McKinneyetal2013} computed using the
M1 method.

\section{Performance}
\label{sec:performance}
\begin{figure}[htp]
\centering
\includegraphics[width=0.9\hsize]{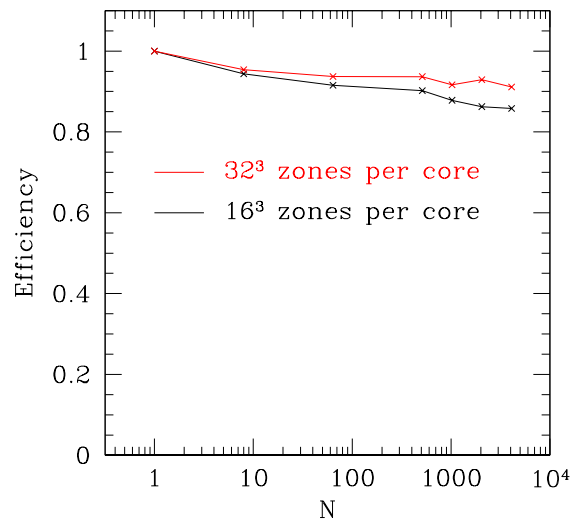}
\caption{Weak scaling of the code measured on {\sc Stampede}, for $32^3$ zones 
per core (red line) and $16^3$ zones per core (black line).
}
\label{scaling}
\end{figure}

The cost to solve the RT equations in our algorithm is linearly proportional to the total number 
of angles in each cell. The time spent in \emph{Step 2}, \emph{Step 4} and \emph{Step 5} 
is roughly the same, when both absorption and scattering opacities are non-zero and the non-linear 
matrix in \emph{Step 4} can converge quickly. To test the performance of the code, we setup a 3D box 
with dimensionless density, pressure, radiation energy density all set to be $1$. We use $80$ angles per 
cell, which are usually enough for the 3D applications we have studied.
The initial intensities are isotropic while the flow velocity is zero initially. The other dimensionless parameters 
are $\Prat=1, \ \Crat=10$. The uniform medium is seeded with $10\%$ random perturbations in density. 
The test is run on {\sc Stampede}, which has Intel Xeon E5 (Sandy Bridge) nodes with core 
frequency $2.7$ GHz and $2$ GB memory per core. With $32^3$ cells per core, the code is able to update 
$N_0=2.4\times 10^4$ zones per second per core. With multiple cores, we define the efficiency to be the 
ratio between the average zones updated by one core per second and $N_0$. The efficiencies of the code 
with total number of cores we use in {\sc Stampede} for two different cases are shown in Figure \ref{scaling}. 
With the number of zones per core fixed to be $32^3$, the parallel efficiency is more than $90\%$ up to 
$4096$ cores. As a comparison, {\sc Athena} without RT can update $\sim1\times10^5$ grid 
cells per second per core for 3D MHD simulations with an efficiency $90\%$ up to $10^5$ processors. Therefore, 
our RT algorithm slows down the code by a factor of $\sim 4$ for $80$ angles per cell 
but keeps a very similar parallel efficiency. This is 
because our explicit RT algorithm does not require iterations in the global domain.  
  
\section{Summary}
\label{sec:summary}
We have developed a new algorithm for radiation MHD based on solving the
\emph{time-dependent} RT equation to compute the specific intensity along discrete
rays.  Integration of the 
specific intensity over angles then yields the radiation energy and momentum source terms, 
which are coupled to the ideal MHD equations.  
We have tested the code extensively based on the suite of test problems used by JSD12.

Compared with commonly used methods for solving the radiation moment equations, such as FLD and M1, the most 
significant advantage of our algorithm is that we calculate the angular distribution of photons self-consistently. 
Therefore, we do not need any ad-hoc closure relation as required in FLD and M1
methods. We expect our methods to be superior in regions where the photon mean free path becomes comparable to or larger 
than the characteristic scales of the fluid. 
Then the disparate contributions of multiple, non-local sources can give rise to complexity in the angular distribution of the photons 
that cannot be encapsulated using only a few of lowest order moments of the radiation field ($E_r$, $\bF_r$).
We have shown
results for a variety of tests that demonstrate the importance of the non-local and anisotropic
properties of the radiation for the correct dynamics, for example the structure
of radiation modified shocks, and the shadows cast by multiple sources of radiation. 

The method described here can be considered an extension of previous algorithms
that use the method of short characteristics to solve the 
\emph{time-independent} RT equation \citep[][]{HayesNorman2003,Davisetal2012}, 
which is then used to calculate a variable Eddington tensor to close
the radiation moment equations \citep[][JSD12]{Stoneetal1992}. 
The full angular distribution of the specific intensity is captured in both algorithms.
However, the main advantages of the 
algorithm described here is that it can be used for dynamical problems where the dynamical time is not negligibly small compared to the light crossing time, 
and when the velocity dependent terms play an important role.  Moreover, since
the method developed here uses explicit differencing of spatial operators, it avoids
the inversion of large matrices every time step, and therefore is much more efficient
on modern parallel systems.  

The algorithm developed in this paper is only applicable to non-relativistic flows.
As general relativity (GR) is believed to be the correct description of the gravitational field near the event horizon of black 
holes, extending our algorithms to GR is the next step, and indeed is already underway.
There are several aspects of the algorithm developed here that we expect will be
crucial for the method in full GR.  This includes the implicit treatment of source
terms, including scattering, and the use of
upwind monotonic interpolation methods that guarantee conservation of 
radiation energy and momentum to roundoff error. 

The time step for stability with our algorithm is based on the light crossing time
of each cell.  This is very inefficient for systems in which the
typical sound speed or flow velocity is much smaller than the speed of light.
In the regime where the reduced speed of light approximation 
applies \citep[][]{SkinnerOstriker2013}, it can be combined with our algorithm to
increase efficiency.
We have used our code to study the 
global structure of black hole accretion disks for different accretion rates at
intermediate distances from the event horizon.  Results from these simulations will
be reported elsewhere.

\section*{Acknowledgements}
 
Support for this work was provided by NASA through 
Einstein Postdoctoral Fellowship grant number 
PF-140109 awarded by the Chandra X-ray Center, 
which is operated by the Smithsonian Astrophysical 
Observatory for NASA under contract NAS8-03060, and by NASA grant NNX11AF49G,
and NSF grant AST-1333612.

\bibliography{FullRT}
\end{CJK*}

\end{document}